\documentclass[twocolumn,showpacs,preprintnumbers,amsmath,amssymb,prb]{revtex4}
\pdfoutput=1
\usepackage{graphicx}
\usepackage{dcolumn}
\usepackage{bm}
\usepackage{braket}

\usepackage[usenames,dvipsnames]{color}
\usepackage{color}
\definecolor{dg}{rgb}{0.00, 0.40, 0.29}

\begin{document}

\title{Competition between excitonic charge and spin density waves:\\ Influence of electron-phonon and Hund's rule couplings}

\author{Tatsuya Kaneko$^1$, Bernd Zenker$^2$, Holger Fehske$^2$, and Yukinori Ohta$^1$}
\affiliation{$^1$Department of Physics, Chiba University, Chiba 263-8522, Japan}
\affiliation{$^2$Institute of Physics, Ernst Moritz Arndt University Greifswald, 17487 Greifswald, Germany}

\date{\today}

\begin{abstract}
We analyze the stability of excitonic ground states in the 
two-band Hubbard  model with additional electron-phonon and Hund's rule 
couplings using a combination of mean-field and variational cluster approaches.  
We show that both the interband Coulomb interaction  and the electron-phonon 
interaction will cooperatively stabilize a charge density wave (CDW) state which typifies 
an ``excitonic'' CDW  if predominantly triggered by the effective interorbital electron-hole attraction or 
a  ``phononic'' CDW  if mostly caused by the coupling to the lattice degrees of freedom. By contrast, 
the  Hund's rule coupling promotes an excitonic spin density wave.  We determine the transition  
between excitonic charge and spin density waves and comment on a fixation of the phase of the excitonic 
order parameter that would prevent the formation of a superfluid condensate of excitons.  The implications
for exciton condensation in several material classes  with strongly correlated electrons are discussed.
\end{abstract} 


\maketitle

\section{Introduction}
Excitonic insulator (EI) phases show a spontaneous coherence between conduction-band electrons and valence-band holes, where the  
prior formation of bound electron-hole pairs (excitons) is typically triggered by the interband Coulomb interaction. The  
condensation of excitons in an EI state was theoretically predicted half a century ago to occur in semiconductors (semimetals) with small band gap (band overlap)\cite{jerome67,halperin68,halperin68-2}. 
The condensed excitonic phase can be characterized by a nonvanishing order parameter  $\langle c^\dag_{\bm{k+Q}}f_{\bm{k}}\rangle$, where $c^\dag_{\bm{k}}$ and  $f^\dag_{\bm{k}}$ are the creation operators of an electron in the conduction and valence bands, respectively.  If the valence-band top and conduction-band bottom are separated  by the wave vector $\bm{Q}$, the system forms a density wave with  modulation $\bm{Q}$.  It is important to note that the EI---despite representing a macroscopic, phase-coherent quantum state---does 
not necessarily feature supertransport properties. 

The experimental efforts to establish the EI in weakly correlated bulk materials  largely failed. 
It is only recently that exciton condensation has been addressed in systems with rather strong electronic correlations\cite{kunes15-2}. 
In this regard, Tm(Se,Te) was argued to exhibit a pressure-induced excitonic instability, related to an anomalous increase in the electrical resistivity and thermal diffusivity~\cite{bucher91,bronold06}.  The charge-density-wave (CDW) state observed in $1T$-TiSe$_2$ was  claimed to 
be of excitonic origin\cite{cercellier07,monney09,monney10,monney11,monney12,zenker13,monney15,watanabe15}.  
In Ca$_{1-x}$La$_x$B$_6$,  the weak ferromagnetism 
was interpreted in terms of  doped spin-triplet excitons\cite{young99,zhitomirsky99,balents00}.   
The condensation of spin-triplet excitons was also predicted to occur in the 
proximity of the spin-state transition,\cite{kunes14} of which 
Pr$_{0.5}$Ca$_{0.5}$CoO$_3$ is an example.\cite{kunes14-2,kunes14-3} 
Likewise, the structural phase transition of the layered chalcogenide 
Ta$_2$NiSe$_5$ has been attributed to a  spin-singlet 
EI\cite{wakisaka09,wakisaka12,kaneko13,seki14}.  
The spin-density-wave (SDW) state of iron-pnictide superconductors 
has sometimes been argued to be of the excitonic origin as 
well\cite{mizokawa08,brydon09,brydon09-2,zocher11}.    
Finally, an EI state was suggested in a $t_{2g}$-orbital system with 
strong spin-orbit coupling\cite{sato15}. 

From the theoretical side, the extended Falicov-Kimball model was considered 
as a paradigmatic model to describe the EI formation and the closely related phenomenon of electronic ferroelectricity\cite{batista02,farkasovsky08,ihle08,zenker10,phan10,zenker10-2,phan11,
zenker11,seki11,zenker12,kaneko13-2,ejima14}.  Here the spin degrees of 
freedom were not taken into account, however.  Excitonic phases in strongly correlated 
spinful systems can be discussed in the framework of two-band Hubbard-type models. 
Including thereby the Hund's rule coupling is known to stabilize the spin-triplet excitonic 
phase in the otherwise degenerate spin-singlet and spin-triplet excitonic 
phases\cite{zocher11,kunes14,kunes14-2,kunes14-3,kaneko12,kaneko14}.   
On the other hand,  we have shown in our previous work\cite{kaneko14} that taking into 
account electronic interactions only, a spin-singlet excitonic phase cannot be stabilized, 
which may however be realized in 1$T$-TiSe$_2$ and Ta$_2$NiSe$_5$, 
where the importance of electron-phonon coupling was recently pointed out\cite{monney11,monney12,zenker13,monney15,watanabe15,kaneko13}.
Although the spin-singlet excitonic state has been investigated in the 
spinless multiband model with electron-phonon coupling\cite{zenker13,phan13,zenker14}, 
not much is known about the role of the electron-phonon coupling played in the 
excitonic density wave states in the spinful multiband Hubbard model.  

Motivated by the recent developments in the field, in this paper,  we will 
thoroughly investigate the stability of the excitonic density wave states in the two-band 
Hubbard model with additional electron-phonon coupling and Hund's rule exchange.  
The model is analyzed employing  static mean-field theory for the electron-phonon coupling and 
the variational cluster approximation (VCA) for the electronic correlations. 
In doing so, we will first show that the interband Coulomb interaction $U'$ and 
electron-phonon interaction $\lambda$ cooperatively stabilize the CDW and that a
smooth crossover occurs between ``excitonic'' CDW and ``phononic'' CDW states, just by 
increasing the ratio $\lambda/U'$. Incorporating the Hund's rule coupling $J$, an excitonic SDW 
state competes with the excitonic CDW.  
The ground-state phase diagram of such extended two-band Hubbard model 
is determined in the  $J$-$\lambda$ plane.  
We will, moreover, pay particular attention to the phase of the order parameter  
in the presence of the electron-phonon and Hund's rule couplings and show 
that both electron-phonon coupling and pair-hopping terms fix the phase 
of the excitonic order parameters, thereby preventing the system from realizing a superfluid.  
Finally, the implications for exciton condensation in real materials will be discussed.  

The paper is organized as follows:  In Sec.~II, we present our model and briefly outline the methods of calculations.  The 
numerical results  will be presented in Sec.~III, where 
electron-phonon interaction,  Hund's rule coupling, and pair-hopping effects will be discussed  and the 
ground-state phase diagram is derived. Section~IV relates our results to recent experiments and draws conclusions.  

\section{Model and Methods}

\subsection{Model Hamiltonian}
We consider the two-band Hubbard model, supplemented by electron-phonon and Hund's rule coupling terms, 
\begin{align}
\mathcal{H}=\mathcal{H}_e+\mathcal{H}^{U}_{e\mathchar`-e} +\mathcal{H}^J_{e\mathchar`-e}
+ \mathcal{H}_{ph} \label{ham}  +\mathcal{H}_{e\mathchar`-ph}\,,
\end{align}
defined on a two-dimensional square lattice. 
The noninteracting $f$/$c$-band electrons are described by 
\begin{align}
\mathcal{H}_e =\sum_{\bm{k},\sigma}\varepsilon^f_{\bm{k}} f^{\dag}_{\bm{k}\sigma}f_{\bm{k}\sigma} 
+  \sum_{\bm{k},\sigma}\varepsilon^c_{\bm{k}} c^{\dag}_{\bm{k}\sigma}c_{\bm{k}\sigma} , 
\end{align}
where $f^{\dag}_{\bm{k}\sigma}$ $(c^{\dag}_{\bm{k}\sigma})$ denotes the 
creation operator of an electron  with momentum $\bm{k}$ 
and spin $\sigma$ $(=\uparrow,\downarrow)$ in the $f$ $(c)$ band.  
Within the tight-binding approximation, the dispersion of band $\alpha$ $(=f,c)$  is given by 
$\varepsilon^{\alpha}_{\bm{k}}=-2t_{\alpha}(\cos k_x + \cos k_y) + \varepsilon_{\alpha}-\mu$, 
where $t_{\alpha}$ is the electron hopping integral between the neighboring sites 
and $\varepsilon_{\alpha}$ is the on-site energy of the $\alpha$ orbital. 
We assume $\varepsilon_f<0$ and $\varepsilon_c>0$, so that the $f$ and $c$ bands correspond to 
the valence and conduction bands, respectively.  The chemical potential  $\mu$ is 
fixed to ensure a filling of two electrons per site (half filling), i.e., 
$\langle n^{f}_i\rangle+\langle n^{c}_i \rangle=2$ with $n^{\alpha}_{i}
=n^{\alpha}_{i\uparrow}+n^{\alpha}_{i\downarrow}
=\alpha^\dag_{i\uparrow}\alpha_{i\uparrow}+\alpha^\dag_{i\downarrow}\alpha_{i\downarrow}$.  

The repulsive Coulomb interaction  takes the form
\begin{align}
\mathcal{H}^{U}_{e\mathchar`-e} = 
U_f\sum_{i}n^f_{i\uparrow}n^f_{i\downarrow} 
+U_c\sum_{i}n^c_{i\uparrow}n^c_{i\downarrow}
+ U' \sum_{i}n^f_{i}n^c_{i},
\end{align}
where $U_{\alpha}$ is its intraorbital part and $U'$ gives the interorbital 
contribution that is responsible for an effective electron-hole attraction and  eventually for an excitonic instability in the system. 
The Hund's exchange interaction is defined by 
\begin{align}
\mathcal{H}^J_{e\mathchar`-e} =&-2J\sum_{i}( \bm{S}^{f}_{i}\cdot\bm{S}^c_{i}+\frac{1}{4}n^f_{i}n^c_{i} ) \notag \\
&-J'\sum_{i}\left( f^{\dag}_{i\uparrow}f^{\dag}_{i\downarrow}c_{i\uparrow}c_{i\downarrow}
+c^{\dag}_{i\uparrow}c^{\dag}_{i\downarrow}f_{i\uparrow}f_{i\downarrow} \right)
\end{align} 
with $\bm{S}^{\alpha}_{i} = \sum_{\sigma,\sigma'}\alpha^{\dag}_{i\sigma}\bm{\sigma}_{\sigma\sigma'}\alpha_{i\sigma'}/2$, 
where $\bm{\sigma}$ is the vector of Pauli matrices.  
$J$ and $J'$ are the strengths of the Hund's rule coupling and 
pair-hopping term, respectively.  $J$ and $J'$ will stabilize a spin-triplet excitonic state\cite{kaneko14}. 

In Eq.~\eqref{ham}, we also included the phonon degrees of freedom because the lattice 
displacements play an important role in the materials under consideration. 
The electron-phonon coupling  becomes particularly important when we address spin-singlet electron-hole 
excitations.  In the harmonic approximation, the phonon part of the  Hamiltonian is given by 
\begin{align}
\mathcal{H}_{ph} = \sum_{\bm{q}} \omega_{\bm{q}} b^{\dag}_{\bm{q}}b_{\bm{q}}\,,
\end{align}
where the bosonic operator $b^{\dag}_{\bm{q}}$ creates a phonon with 
momentum $\bm{q}$ and frequency $\omega_{\bm{q}}$   
(we have set $\hbar=1$).  The dominant electron-phonon coupling term between a $c$-$f$ (electron-hole) 
excitation and lattice displacement is assumed to be  
\begin{align}
\mathcal{H}_{e\mathchar`-ph}= \frac{1}{\sqrt{N}}\sum_{\bm{k},\bm{q}}\sum_{\sigma}g_{\bm{q}} 
(b_{\bm{q}}+b^{\dag}_{-\bm{q}}) c^{\dag}_{\bm{k}+\bm{q}\sigma}f_{\bm{k}\sigma} + \mathrm{H.c.}
\label{hamep}
\end{align}
with  coupling 
constant $g_{\bm{q}}$ \cite{zenker13,watanabe15,phan13,zenker14}.

Throughout the paper, we fix the hopping parameters $t_f=t_c=t$ and use $t$ as the unit of energy.  
Furthermore, we set $\varepsilon_c/t=-\varepsilon_f/t=3.2$, so that the noninteracting 
band structure represents a semimetal with a small band overlap.  
The conduction-band bottom at $\bm{k}=(0,0)$ gives rise to an electron pocket, while the valence-band top produces a hole pocket at 
$\bm{k}=(\pi,\pi)$, resulting  in the modulation vector of the density wave  $\bm{Q}=(\pi,\pi)$; see Ref.~\onlinecite{kaneko12} 
for the band dispersion and Fermi surface in the Brillouin zone of the square lattice.
For simplicity, we assume $U_f=U_c=U$ and employ $U=2U'-J$ to 
suppress the Hartree shift\cite{al}. In this choice, the EI state is stabilized between the band-insulator and 
Mott-insulator states\cite{kaneko12,kaneko14}.   
Moreover, we consider a dispersionless Einstein phonon $\omega_{\bm{q}}=\omega$ and 
a momentum-independent electron-phonon coupling constant $g_{\bm{q}}=g$.  
Since the strength of the electron-phonon coupling appears in the form 
$\lambda = g^2/\omega$ in the mean-field approximation used below, 
we take $\lambda$ as the electron-phonon coupling parameter in what follows. 

\subsection{Mean-field approximation for the phonons}
We treat the electron-phonon interaction term $\mathcal{H}_{e\mathchar`-ph}$ 
in the mean-field (frozen-phonon) approximation.  
Introducing the expectation values of the $c$-$f$ hybridization 
$\langle c^{\dag}f\rangle$ and lattice displacement 
$\langle b \rangle$, the operators in Eq.~(\ref{hamep}) are 
approximated as 
$b_{\bm{q}} c^{\dag}_{\bm{k}+\bm{q}\sigma}f_{\bm{k}\sigma} 
\sim  [\, \langle b_{\bm{q}} \rangle  c^{\dag}_{\bm{k}
+\bm{q}\sigma} f_{\bm{k}\sigma} +b_{\bm{q}} 
\langle c^{\dag}_{\bm{k}+\bm{q}\sigma} f_{\bm{k}\sigma} \rangle \, ] 
\delta_{\bm{q},\bm{Q}} -\langle b_{\bm{q}} \rangle 
\langle c^{\dag}_{\bm{k}+\bm{q}\sigma} f_{\bm{k}\sigma} 
\rangle  \delta_{\bm{q},\bm{Q}}$.  
Since in our model the nesting vector $\bm{Q}=(\pi,\pi)$ is commensurate with the lattice 
periodicity, $e^{2i\bm{Q}\cdot\bm{r_i}}=1$ for lattice vectors $\bm{r}_i$. 
Hence, we have 
$b_{\bm{Q}} = b_{-\bm{Q}}$ ($b^{\dag}_{\bm{Q}} = b^{\dag}_{-\bm{Q}}$) \cite{zenker13}, 
where $b_{\bm{Q}}$ and $b_{-\bm{Q}}$ ($b^{\dag}_{\bm{Q}}$ and $b^{\dag}_{-\bm{Q}}$) 
annihilate (create) the same phonon.  
This implies $\langle b_{\bm{Q}} \rangle = \langle b_{-\bm{Q}} \rangle = \langle b^{\dag}_{\bm{Q}}\rangle$, 
and therefore $\langle b_{\bm{Q}} \rangle$ becomes a real number.  
In view of $\langle c^{\dag}_{\bm{k}+\bm{Q}\sigma} f_{\bm{k}\sigma} \rangle 
=\langle  f^{\dag}_{\bm{k}\sigma} c_{\bm{k}+\bm{Q}\sigma} \rangle^{*} \ne 0$, 
we define the complex order parameter of the excitonic CDW  as 
\begin{align}
\Phi_c = |\Phi_c| e^{i\theta_c} = \frac{1}{2N} \sum_{\bm{k},\sigma} 
\langle c^{\dag}_{\bm{k}+\bm{Q}\sigma} f_{\bm{k}\sigma} \rangle , \label{CDWOP}
\end{align} 
where $|\Phi_c|$ and $\theta_c$ are the amplitude and phase of 
the order parameter, respectively.  Then the electron-phonon part
in the mean-field approximation  is 
\begin{align}
&\mathcal{H}^{\mathrm{MF}}_{e\mathchar`-ph}
= \frac{2g}{\sqrt{N}}\langle b_{\bm{Q}} \rangle \sum_{\bm{k},\sigma} 
\left( c^{\dag}_{\bm{k}+\bm{Q}\sigma} f_{\bm{k}\sigma} +  f^{\dag}_{\bm{k}\sigma} 
c_{\bm{k}+\bm{Q}\sigma} \right) \notag \\
&+ 4g\sqrt{N} (b^{\dag}_{\bm{Q}}+b_{\bm{Q}}) |\Phi_c|\cos \theta_c 
-8g\sqrt{N}\langle b_{\bm{Q}} \rangle |\Phi_c| \cos \theta_c . \label{MFep}
\end{align}
Introducing $B_{\bm{q}} = b_{\bm{q}} + \delta_{\bm{q},\bm{Q}}(4g\sqrt{N}/\omega) |\Phi_c|\cos \theta_c$, 
the phonon Hamiltonian $\mathcal{H}_{ph}$ together with the second term of 
the right-hand side of Eq.~(\ref{MFep}) can be diagonalized, yielding  
$\omega \sum_{\bm{q}} B^{\dag}_{\bm{q}}B_{\bm{q}} - 16\lambda N |\Phi_c|^2\cos^2\theta_c$.  
Hence, from $\langle B_{\bm{Q}} \rangle = \langle B^{\dag}_{\bm{Q}} \rangle=0$, 
we find 
\begin{align}
\langle b_{\bm{Q}} \rangle = \langle b^{\dag}_{\bm{Q}} \rangle = -\frac{4g\sqrt{N}}{\omega}|\Phi_c | \cos \theta_c . \label{bQ}
\end{align}
Substituting this expression into Eq.~(\ref{MFep}), 
we finally obtain the mean-field electron-phonon Hamiltonian, 
\begin{align}
\mathcal{H}^{\mathrm{MF}}_{e\mathchar`-ph}
=& \Delta_p \cos \theta_c \sum_{\bm{k},\sigma}  
f^{\dag}_{\bm{k}\sigma} c_{\bm{k}+\bm{Q}\sigma}  + \mathrm{H.c.}  
+ \frac{N\Delta_p^2}{4\lambda}  \cos ^2 \theta_c  \label{MFep2}
\end{align}
with  $\Delta_p = -8\lambda |\Phi_c|$.  
Using Eq.~(\ref{MFep2}), below we will minimize the grand potential of the system with respect to $\Delta_p$ and $\theta_c$.

We define the complex order parameter of the excitonic SDW as 
\begin{align}
\Phi_s = |\Phi_s|e^{i\theta_s} = \frac{1}{2N}\sum_{\bm{k},\sigma}\sigma
\langle c^{\dag}_{\bm{k}+\bm{Q}\sigma}f_{\bm{k}\sigma} \rangle , \label{SDWOP} 
\end{align}
where $|\Phi_s|$ and $\theta_s$ are the amplitude and phase of the order parameter, respectively.  
Because we assume an SDW state with modulation vector $\bm{Q}=(\pi,\pi)$, where 
the expectation value $\langle c^{\dag}_{i\uparrow}f_{i\uparrow} \rangle$ is in antiphase compared to 
$\langle c^{\dag}_{i\downarrow}f_{i\downarrow} \rangle$ regarding the spatial variation, 
these two expectation values have opposite signs on the same site.  
In momentum space, this reads $\sum_{\bm{k}}\langle c^{\dag}_{\bm{k}+\bm{Q}\uparrow}f_{\bm{k}\uparrow} \rangle
=-\sum_{\bm{k}}\langle c^{\dag}_{\bm{k}+\bm{Q}\downarrow}f_{\bm{k}\downarrow} \rangle$.  
We then find, from Eqs.~(\ref{CDWOP}) and (\ref{bQ}), that 
$\langle b_{\bm{Q}}\rangle = \langle b^{\dag}_{\bm{Q}} \rangle \propto 
\sum_{\bm{k},\sigma} \langle c^{\dag}_{\bm{k}+\bm{Q}\sigma} f_{\bm{k}\sigma} \rangle=0$, 
which means that the spin-triplet condensate will not couple to the  phonons.  

\subsection{Variational cluster approximation}
In order to take electron correlation effects into account, we treat the electronic interactions in \eqref{ham} within the VCA,\cite{potthoff03,dahnken04} 
which is a  quantum cluster method based on the  self-energy functional theory\cite{potthoff03-2}.  
The VCA first introduces disconnected clusters of finite size, for which the 
cluster self-energy $\Sigma '$  can be computed exactly. In a next step, out of this,  a superlattice is formed as a reference system.  
By restricting the  trial self-energy to $\Sigma '$, we obtain a 
certain approximation to the grand potential of the original system, 
\begin{align}
\Omega=\Omega'+\mathrm{Tr}\: 
\mathrm{ln} ( G^{-1}_0-\Sigma' )^{-1} - \mathrm{Tr}\: \mathrm{ln}( G' ), \label{gp}
\end{align}
where $\Omega'$ and $G'$ are the grand potential and Green's function 
of the reference system, respectively, and $G_0$ is the noninteracting 
Green's function; for further details, see Refs.~\onlinecite{potthoff12,senechal08}.
In doing so, the short-range electron correlations within the cluster of the reference 
system are treated exactly. 
In our VCA calculation, we take an $L_c=2\times2=4$ site (eight-orbital) cluster 
as the reference system and we use exact diagonalization to solve 
the corresponding quantum many-body problem in the cluster. 
Within VCA, we can  take into account spontaneous symmetry breakings just by adding 
appropriate Weiss fields to the reference system\cite{dahnken04}, and take these fields as variational parameters. 
The Weiss fields for 
excitonic CDW and SDW states, which are defined by the order parameter 
$\Phi_c$ [in Eq.~(\ref{CDWOP})] and $\Phi_s$ [in Eq.~(\ref{SDWOP})], 
respectively, may be written as 
\begin{align}
&\mathcal{H}^{\mathrm{WF}}_{\mathrm{c}}=  \Delta'_0 e^{i\theta_c} \sum_{\bm{k},\sigma} f^{\dag}_{\bm{k}\sigma}
c_{\bm{k}+\bm{Q}\sigma} + \mathrm{H.c.}  \\
&\mathcal{H}^{\mathrm{WF}}_{\mathrm{s}}=  \Delta'_s e^{i\theta_s} \sum_{\bm{k},\sigma} \sigma f^{\dag}_{\bm{k}\sigma}
c_{\bm{k}+\bm{Q}\sigma} + \mathrm{H.c.}\,.
\end{align}
Here, $\Delta'_0$ and $\Delta'_s$ are the strengths of the Weiss fields 
for the excitonic CDW and SDW states generated by  $\mathcal{H}^{U}_{e\mathchar`-e}$ and $\mathcal{H}^J_{e\mathchar`-e}$. 

According to Eq.~\eqref{MFep2} , we take into account the contribution of the phonons in the mean-field approximation  as a one-particle term in the original system.  
Then, the Hamiltonian describing an excitonic CDW state in the reference system is given by 
\begin{align}
\mathcal{H}'_{\mathrm{c}}=\mathcal{H}_e+\mathcal{H}^{U}_{e\mathchar`-e} +\mathcal{H}^J_{e\mathchar`-e} 
+ \mathcal{H}^{\mathrm{MF}}_{e\mathchar`-ph} + \mathcal{H}^{\mathrm{WF}}_{\mathrm{c}}, 
\end{align}
where we note that $\mathcal{H}_e+\mathcal{H}^{U}_{e\mathchar`-e} +\mathcal{H}^J_{e\mathchar`-e} + \mathcal{H}^{\mathrm{MF}}_{e\mathchar`-ph}$ 
is the Hamiltonian of the original system and the Weiss field $\mathcal{H}^{\mathrm{WF}}_{\mathrm{c}}$ is added in the reference system.  
Using $\mathcal{H}'_{\mathrm{c}}$, we calculate the grand potential $\Omega$  and 
optimize the variational parameters $\Delta'_0$, $\Delta_p$, and $\theta_c$.  
The most stable solution with $(\Delta_0',\Delta_p)\ne (0,0)$ corresponds to the excitonic CDW state. 
Note that we determine the parameters $\Delta_p$ and $\theta_c$ via the minimization of the grand potential 
rather than solving the self-consistent equation.  
Both procedures are equivalent, however, since the order 
parameter $\Phi_c$ calculated, using the Green's function with $\Delta_p$ and $\theta_c$ optimized via the grand potential calculation in VCA, exactly satisfies the self-consistent condition 
$\Delta_p= 8\lambda \Phi_c$.  

Since the spin-triplet term does not couple to the lattice degrees of freedom within our mean-field approach, the phonons 
will not affect the excitonic SDW state.  
Then the Hamiltonian of the reference system describing an excitonic SDW is 
\begin{align}
\mathcal{H}'_{\mathrm{s}}=\mathcal{H}_e+\mathcal{H}^{U}_{e\mathchar`-e} +\mathcal{H}^J_{e\mathchar`-e}
+ \mathcal{H}^{\mathrm{WF}}_{\mathrm{s}}.
\end{align}
Again we calculate the grand potential $\Omega$ from the reference Hamiltonian 
$\mathcal{H}'_{\mathrm{s}}$ and optimize variational parameters $\Delta'_s$ and $\theta_s$, where the most stable solution with $\Delta_s' \ne 0$ corresponds to the SDW state.

\section{Numerical results}

\subsection{Phase of the order parameters}
\begin{figure}[!tb]
\begin{center}
\includegraphics[width=\columnwidth]{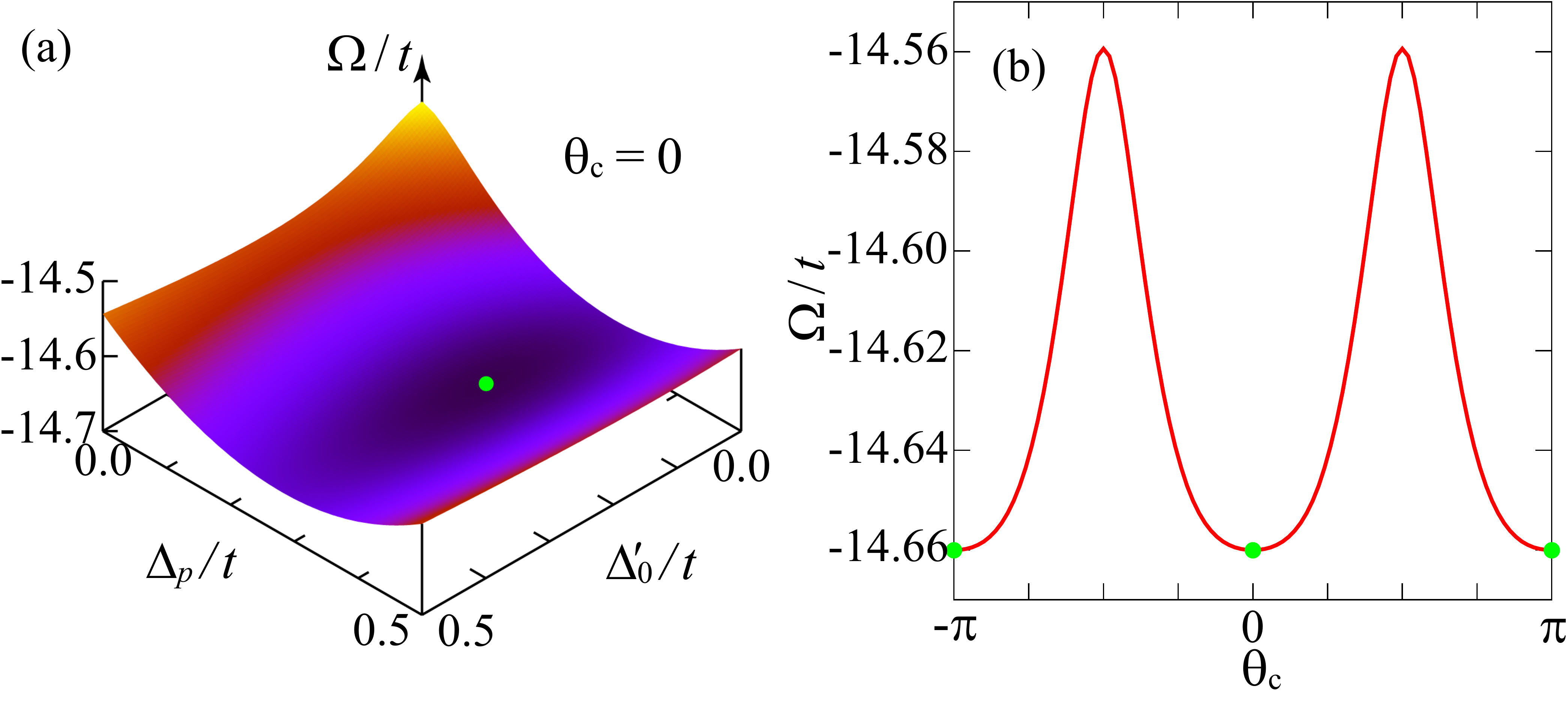}
\caption{(Color online) 
(a)  Grand potential $\Omega$ as a function of the variational 
parameters $\Delta_0'$ and $\Delta_p$ for $U'/t=4$ and $\lambda/t=0.15$.  
(b) $\theta_c$ dependence of $\Omega$ taken  the values of $\Delta_0'$ and $\Delta_p$ 
optimized at $\theta_c=0$.  Green dots indicate the  stationary points. 
}\label{fig1}
\end{center}
\end{figure}

We first discuss the phase of the different order parameters entering the grand potential.
In the spin-singlet excitonic state, the system forms an  
excitonic CDW at any finite $U'$ and $\lambda$   due to the perfect nesting 
of the Fermi surface.  Figure~\ref{fig1}(a) shows 
the calculated grand potential $\Omega$ as a function of 
the variational parameters $\Delta_0'$  and $\Delta_p$.  
Obviously the grand potential has a stationary point at 
$(\Delta_0',\Delta_p)\ne(0,0)$, signaling a CDW ordering.  
Without electron-phonon coupling, $\Omega$ is independent of the phase 
$\theta_c$, i.e., $\Omega(\theta_c)=\Omega(\theta'_c)$.  
Accordingly, the excitonic CDW state reveals a gapless acoustic phase mode 
in its excitation spectrum\cite{zenker14}. 
If, however, the electron-phonon coupling comes into play, 
the grand potential manifests a dependence on the phase of the (complex) order parameter.  
In Fig.~\ref{fig1}(b), we display the  $\theta_c$ dependence of $\Omega$; 
the grand potential takes its minimum at $\theta_c = 0, \pi$. This phase fixation 
may be expected looking at Eq.~(\ref{MFep2}).  In our mean-field approximation, 
the single-particle gap caused by $\lambda$ is given as $\Delta_p \cos \theta_c$ and  
is maximized at $\theta_c = 0$.  When $\theta_c$ is fixed by the electron-phonon coupling, 
the collective phase mode in the spin-singlet excitonic state becomes massive
(see the discussion of the spinless model in 
Ref.~\onlinecite{zenker14}).

\begin{figure}[!tb]
\begin{center}
\includegraphics[width=\columnwidth]{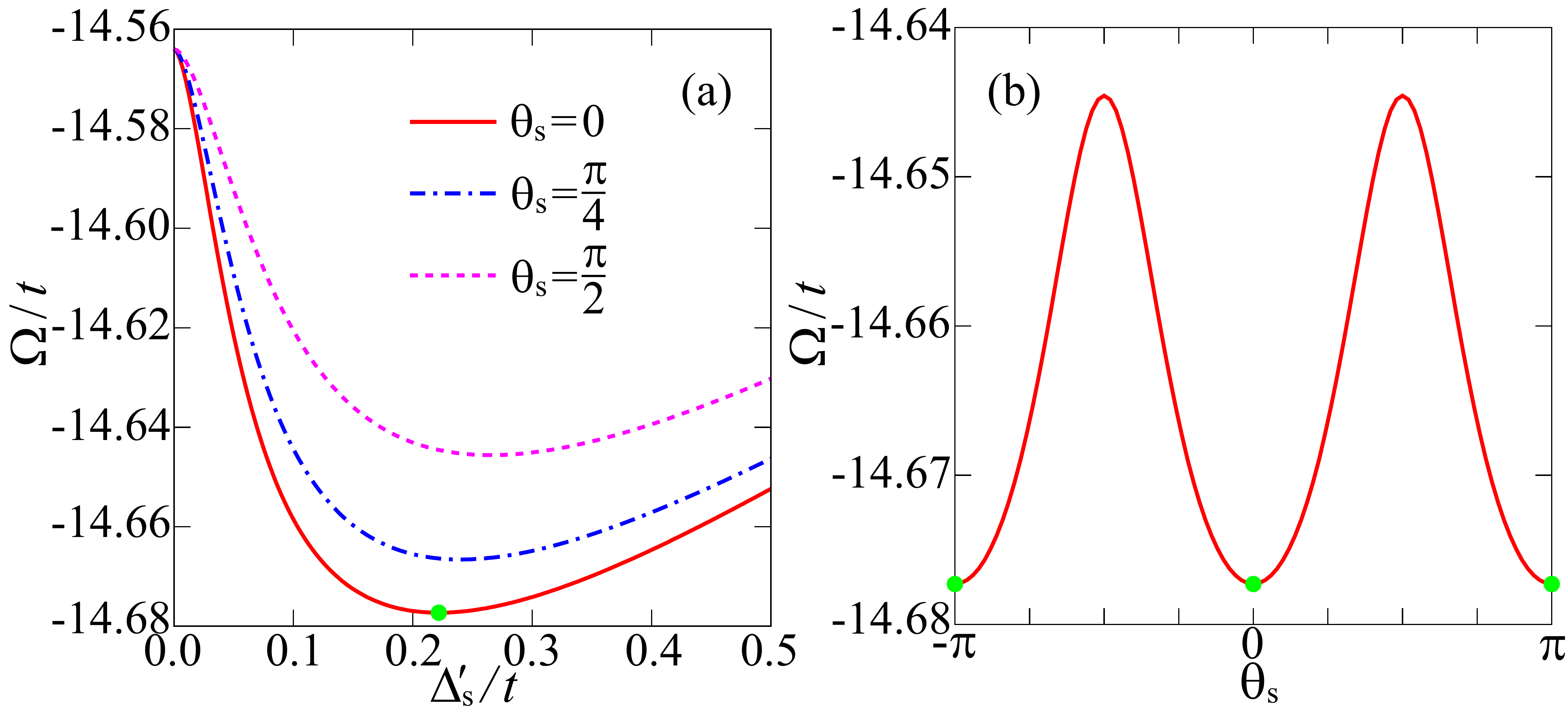}
\caption{(Color online) 
(a) Grand potential $\Omega$ as a function of the variational 
parameter $\Delta_s'$ at $U'/t=4$ and $J/t=J'/t=1$. 
(b)  $\theta_s$ dependence of  $\Omega$ obtained using the value of $\Delta_s'$ optimized at $\theta_s=0$.  
Dots mark stable stationary points.  
}\label{fig2}
\end{center}
\end{figure}

In the case of the spin-triplet excitonic state, the excitonic SDW and CDW states are degenerate 
if the electron-phonon and Hund's couplings are neglected.  The Hund's exchange terms $\propto J$ 
and $\propto J'$ lift this degeneracy and stabilize the SDW state\cite{kaneko14}.  
Note that the $\theta_s$ dependence of the grand potential 
behaves differently in the presence or absence of the pair-hopping term $J'$: 
For $J'=0$, the grand potential of the SDW state does not depend 
on $\theta_s$, i.e., $\Omega(\theta_s)=\Omega(\theta'_s)$,  
whereas $\Omega$ depends on $\theta_s$ at any finite $J'$.  
Again the independence of $\Phi_s$ on the phase value $\theta_s$ accounts for a gapless 
excitation spectrum, i.e., an acoustic phase mode.  Figure~\ref{fig2} gives 
the calculated grand potential $\Omega$ as a function of the phase $\theta_s$ 
in the presence of the pair-hopping term $J'$. Indeed we  find that $\Omega$ has 
two minima, at $\theta_s = 0, \pi$, which fixes the phase $\theta_s$ 
of $\Phi_s$. It is known that the energy in the presence of the pair-hopping-type 
exchange interaction shows a phase dependence  
$\cos 2\theta_s$\cite{littlewood96}. This is why the pair-hopping term $J'$ fixes $\theta_s$
and in that way destroys the gapless acoustic phase mode in the spin-triplet excitonic state.  

\begin{figure}[tb]
\begin{center}
\includegraphics[width=0.95\columnwidth]{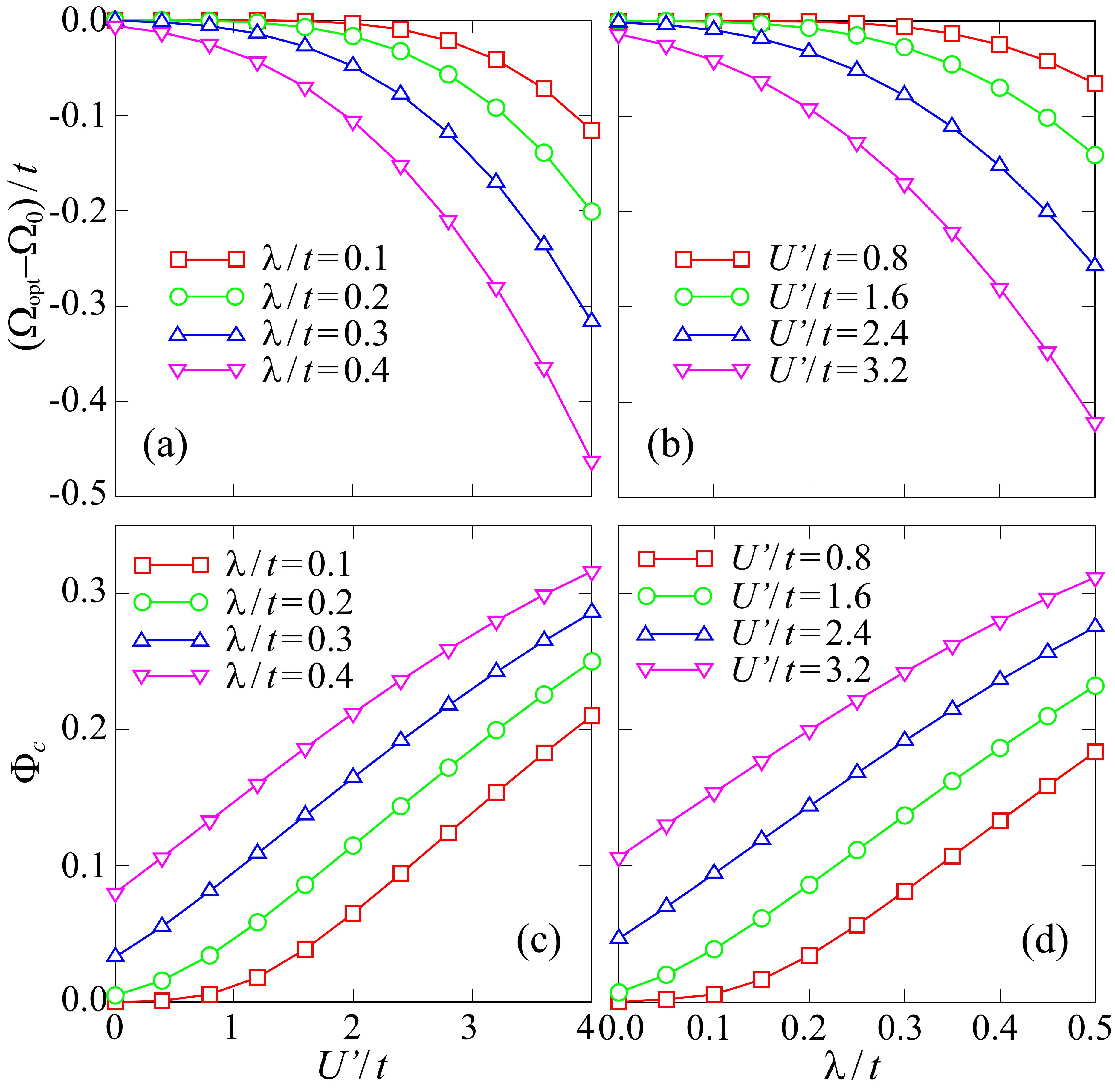}
\caption{(Color online) 
Optimized values of the grand potential $\Omega_\mathrm{opt}$ 
in dependence on  (a) $U'/t$  and (b) $\lambda/t$.   Here, $\Omega_0$ is the grand potential 
in the normal (semimetallic) state. Order parameter $\Phi_c$ for the excitonic CDW state as a function of 
(c) $U'/t$ and (d) $\lambda/t$.    
}\label{fig3}
\end{center}
\end{figure}

\subsection{Excitonic CDW state}
Now let us analyze the stability of the CDW state in the presence of the electron-phonon coupling in more detail.  
In Fig.~\ref{fig3}, we present the results for both the optimized grand potential $\Omega_{\mathrm{opt}}$ and the order parameter $\Phi_c$ when  the interband Coulomb interaction $U'$ and the electron-phonon coupling $\lambda$ are varied.  
$\Omega_{\mathrm{opt}}$ indicates that (i) the symmetry-broken CDW state is lower in energy than the normal state and (ii) the stability of the CDW state is enhanced if $U'$ and $\lambda$ are increased; see Figs.~\ref{fig3}(a) and \ref{fig3}(b).  
This is corroborated by the behavior of the order parameter $\Phi_c$ displayed in Figs.~\ref{fig3}(c) and \ref{fig3}(d).
We see that the interband Coulomb interaction $U'$ induces and boosts the excitonic CDW state while the electron-phonon coupling $\lambda$ rather promotes a phononic CDW state (see below). 
Both, however, cooperatively stabilize a charge-ordered state.  
In this connection, the electron-phonon coupling  lifts the degeneracy of CDW and SDW that exists for $\lambda=0$.  

\begin{figure}[tb]
\begin{center}
\includegraphics[width=0.95\columnwidth]{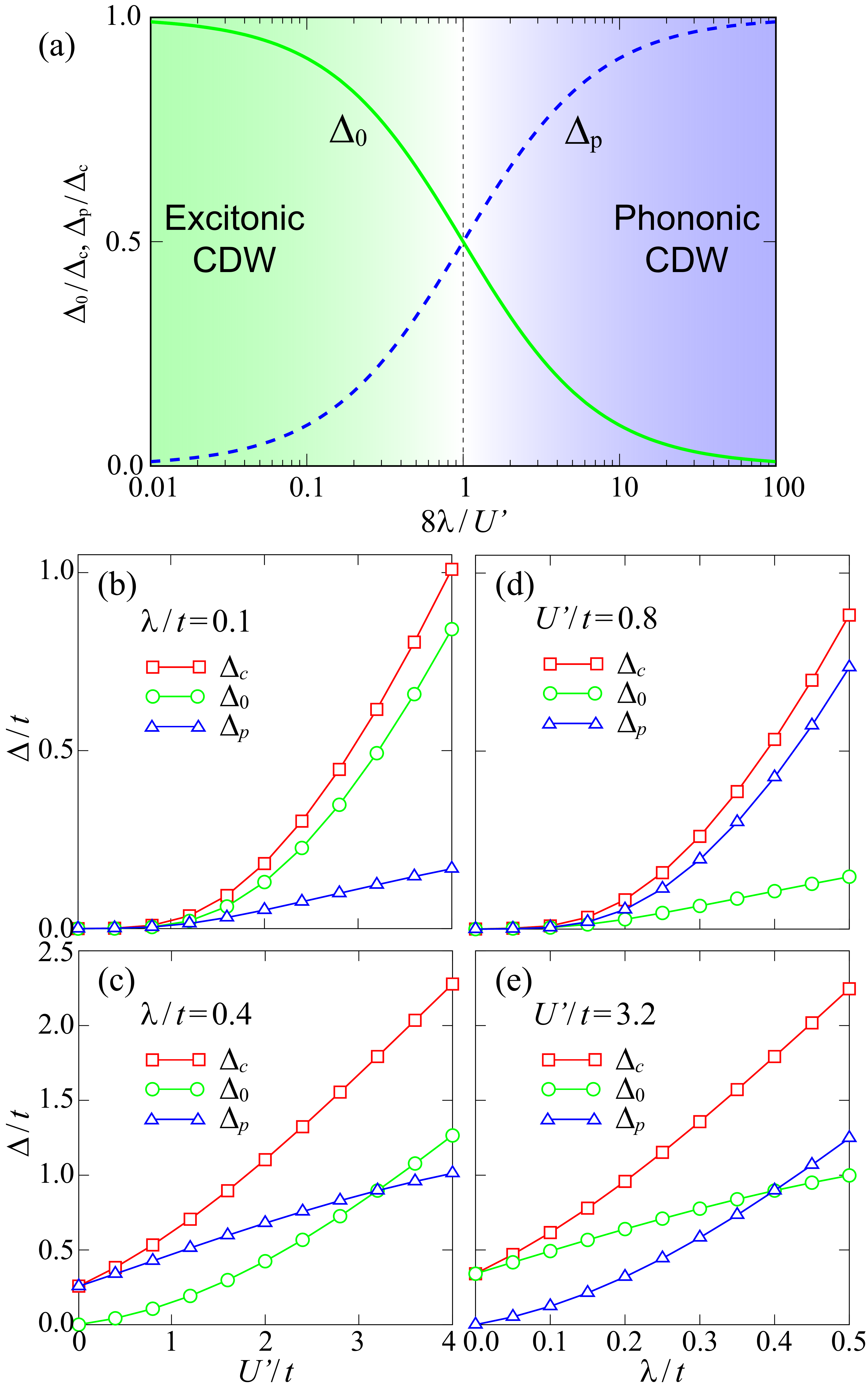}
\caption{(Color online) 
(a) Excitonic gap parameter $\Delta_0$ (solid line) and phononic 
gap parameter $\Delta_p$ (dashed line) divided by the total gap 
$\Delta_c=\Delta_0+\Delta_p$.   $\Delta_c$, $\Delta_0$, and $\Delta_p$ 
are separately plotted as a function of (b),(c) $U'/t$ and (d),(e) $\lambda/t$.  
}\label{fig4}
\end{center}
\end{figure}

In the mean-field  approximation, the gap parameter of the CDW state, $\Delta_c=(U'+8\lambda)\Phi_c$, can be separated into two contributions:  the excitonic (or interband Coulomb driven) part $\Delta_0=U'\Phi_c$ 
and the phononic (or electron-phonon driven) part $\Delta_p=8\lambda \Phi_c$.  
Figure~\ref{fig4}(a) illustrates the relative magnitude of  $\Delta_0$ and $\Delta_p$, in dependence 
on the ratio $8\lambda/U'$.  At $8\lambda / U' \ll 1$, $\Delta_c \simeq \Delta_0 \gg \Delta_p$ and 
the CDW state, stabilized by the interband Coulomb interaction $U'$, is excitonic by its nature. 
Increasing $8\lambda / U'$, $\Delta_0$ decreases while $\Delta_p$ 
increases, indicating a smooth crossover to a phononic CDW, which  fully develops 
at $8\lambda / U' \gg 1$, where $\Delta_c \simeq \Delta_p \gg \Delta_0$. In the crossover 
region  $8\lambda/U'\simeq 1$, both excitonic and phononic contributions are equally important.

In Figs.~\ref{fig4}(b)--\ref{fig4}(e), we show the behavior of the different contributions to the gap 
parameter $\Delta_c$ when $U'$ and $\lambda$ are varied separately. Data are obtained by VCA.
Enhancing $U'/t$ ($\lambda/t$) at weak $\lambda/t$ (small $U'/t$) leads to an increase
in $\Delta_p$ ($\Delta_0$) as well, since both interactions couple to the same operator-product
expectation value $\langle c^{\dag}_{\bm{k}+\bm{Q}\sigma} f_{\bm{k}\sigma} \rangle$; see Figs.~\ref{fig4}(b)
and \ref{fig4}(d). The crossover between excitonic and phononic CDWs can be seen in Figs.~\ref{fig4}(c)
and \ref{fig4}(e), where a crossing between $\Delta_p$ and $\Delta_0$ appears when $U'\simeq 8\lambda$.

\begin{figure}[t]
\begin{center}
\includegraphics[width=0.95\columnwidth]{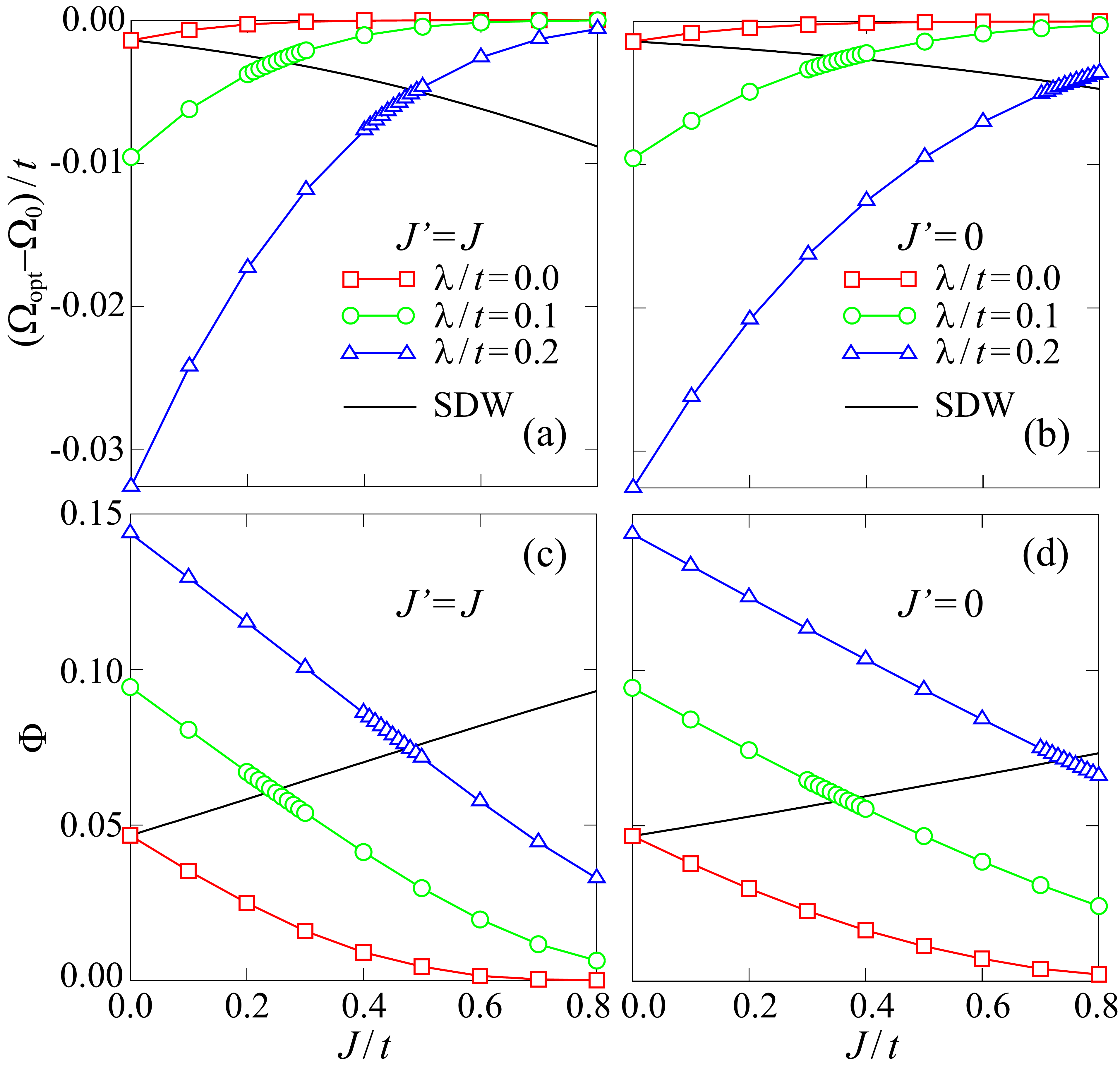}
\caption{(Color online) 
$J$ dependence of the grand potential $\Omega_\mathrm{opt}$ 
and the order parameter $\Phi$ in the excitonic CDW (symbols) and SDW (solid line) states (a),(c) 
with  ($J'=J$) and (b),(d) without ($J'=0$) 
 the pair-hopping term, where $U'/t=2.4$. $\Omega_0$ is the grand 
potential of the normal semimetallic state. 
}\label{fig5}
\end{center}
\end{figure}
 
\begin{figure}[h]
\begin{center}
\includegraphics[width=0.9\columnwidth]{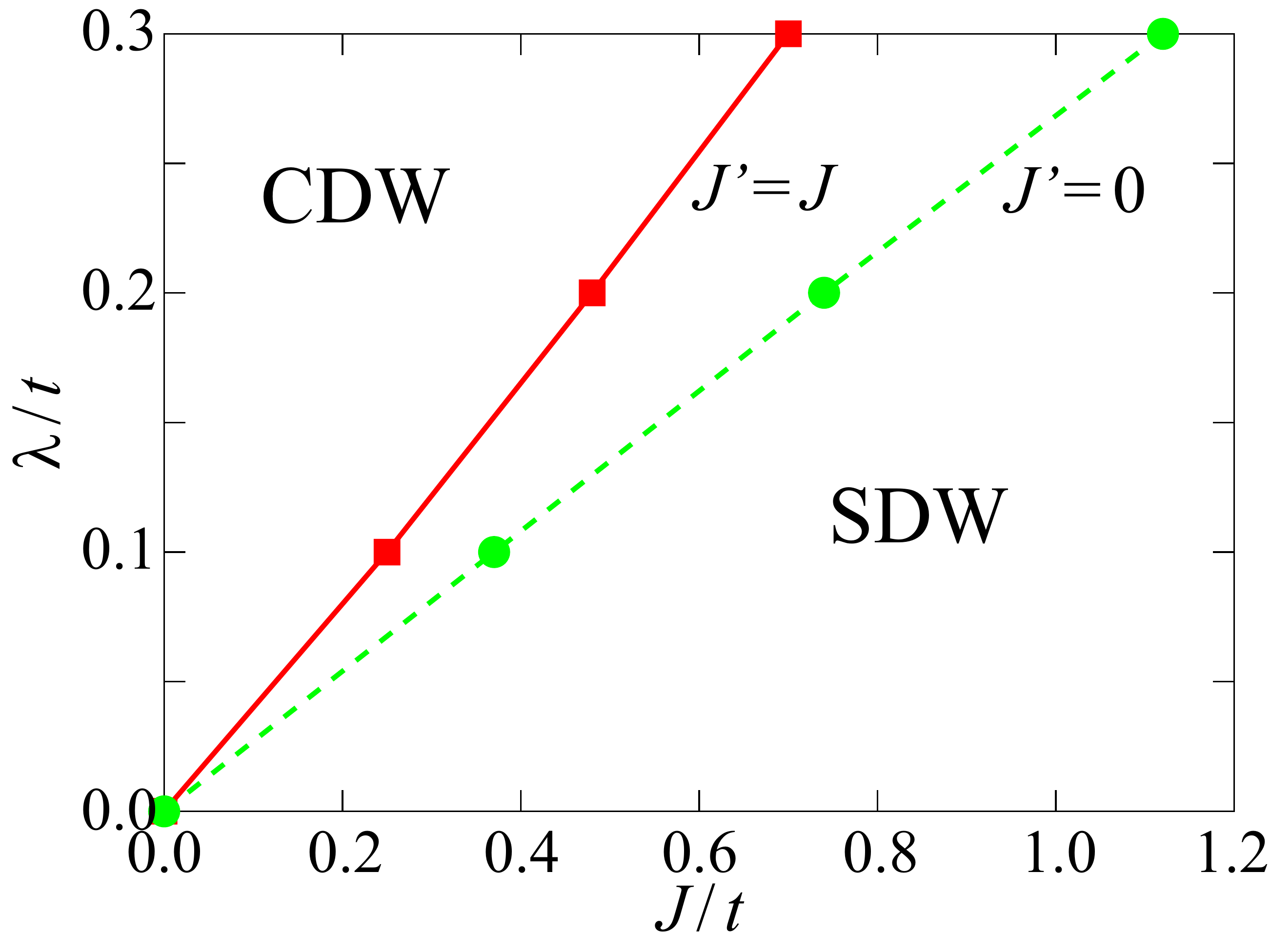}
\caption{(Color online) 
Ground-state phase diagram of the two-band Hubbard model with electron-phonon and Hund's rule couplings showing the stability regions of excitonic CDW and SDW phases. Results are obtained, in the presence ($J'=J$) and absence ($J'=0$) of the pair-hopping term, by combining the mean-field and VCA approaches, for a two-dimensional (square) lattice at half filling, where $U'/t=2.4$.} 
\label{fig6}
\end{center}
\end{figure}

\subsection{Excitonic SDW state}
We now study the influence of the Hund's rule coupling on the nature of the excitonic phase,
and also when an additional electron-phonon coupling acts in the system.  
Evidently, excitonic CDW and SDW states are degenerate 
at $J=J'=0$ and $\lambda=0$~\cite{kaneko12,kaneko14}.
Any finite $J$ and/or $\lambda$ lifts this degeneracy. 
Figure~\ref{fig5} clearly shows that by increasing $J$, the optimized 
grand potential $\Omega_\mathrm{opt}$ for the SDW (CDW) state monotonically 
decreases (increases); accordingly, the order parameter for the SDW (CDW) phase
is enhanced (suppressed). This holds for both $J'>0$ and $J'=0$. Clearly the SDW state is stable as soon as 
$\Omega_\mathrm{opt}^\mathrm{SDW}$ becomes less than $\Omega_\mathrm{opt}^\mathrm{CDW}$.
A finite  pair-hopping term $\propto J'$ amplifies the tendency towards SDW formation\cite{kaneko14}.  
 
\subsection{Ground-state phase diagram}
The competition between electron-phonon and Hund's rule coupling effects leads to the ground-state
phase diagram of the model \eqref{ham} presented in Fig.~\ref{fig6}. Obviously,  $\lambda$ and $J$
tend to establish CDW and SDW phases, respectively, on top of an excitonic state enforced by $U'$.  A finite $J'$
increases the region in the $J$-$\lambda$ plane where the excitonic SDW is the ground state.
We note that the SDW-CDW transition is a first-order transition, within the limits of our approximations.

\section{Discussion and conclusions}
First, let us discuss implications of our findings on materials aspects.  
The transition-metal chalcogenides 1$T$-TiSe$_2$ and Ta$_2$NiSe$_5$ have 
recently been discussed in terms of the spin-singlet EI.  
In these systems, the valence and conduction bands are formed by orbitals located on different atoms.
For example, in $1T$-TiSe$_2$, the $4p$ orbitals of Se ions account for the valence 
bands and the $3d$ orbitals of Ti ions account for the conduction 
bands\cite{cercellier07,monney09,monney10,monney11,monney12,zenker13,monney15,watanabe15}, 
and in Ta$_2$NiSe$_5$, the $3d$ orbitals of Ni ions form the valence 
bands and the $5d$ orbitals of Ta ions form the conduction 
bands\cite{wakisaka09,wakisaka12,kaneko13,seki14}. 
Hund's rule coupling, acting between electrons on different 
orbitals of a single ion and favoring the spin-triplet excitons, 
is therefore negligible.  Rather, in these materials, 
the electron-phonon coupling is at play and will stabilize a spin-singlet 
EI state.  The interband Coulomb interaction and 
electron-phonon interaction, which are inherently interrelated  
in these materials, will cooperatively stabilize the EI CDW, which 
is predominantly phononic or excitonic depending on the importance of electron-phonon or Coulomb effects. 

By contrast, in the iron-pnictide superconductors\cite{brydon09,brydon09-2,zocher11} and Co oxides\cite{kunes14,kunes14-2,kunes14-3}, the valence and conduction bands are formed by the $d$ orbitals 
on the (same) transition-metal ions, so that the Hund's rule coupling is expected to be strong. 
Hence, in these materials, the SDW phase, if really excitonic in origin, is rather triggered by the Hund's rule coupling than by electron-phonon coupling. Then, as our phase diagram suggests, the condensation of spin-triplet excitons will play a major role.

Second, let us comment on the phase of the excitonic order 
parameters. On the one hand, as we have shown in the preceding section, 
the electron-phonon interaction stabilizes the spin-singlet excitonic condensate,  
whereas exchange interactions such as the Hund's rule couplings stabilize 
a spin-triplet excitonic condensate in the otherwise degenerate 
excitonic density-wave states.  On the other hand, these interactions, in 
particular the electron-phonon and pair-hopping interactions, will fix the phase of 
the order parameter of the excitonic state; see Sec.~III~A.   
Because the spatial modulations of the CDW and SDW  
are given by $\cos(\bm{Q}\cdot\bm{r}_i + \theta)$, the phase $\theta$ 
may lead to a translational motion of the condensate as a whole\cite{gruner}. 
If the energy of the condensate is independent of the phase, maintaining  
the continuous symmetry of the system with respect to the phase, 
a gapless acoustic phase mode may appear in the 
excitation spectrum, allowing for a  translational motion of the condensate without loss of energy 
(i.e., superfluidity), as predicted by Fr\"ohlich in his theory of 
incommensurate density waves\cite{frohlich54}.  
In real materials, however, excitonic condensation will be influenced by the 
lattice degrees of freedom or affected by the pair-hopping term. 
Then the phase of the condensate is fixed and a gap opens 
for the collective phase mode.  This makes realization of excitonic superfluidity in real materials 
unlikely. 

To summarize, we have studied the stability of the excitonic states with charge and 
spin density  modulations  in terms of the two-band Hubbard model, supplemented by 
electron-phonon and Hund's rule interactions, where the static mean-field 
theory is employed for coupling to the lattice degrees of freedom and the 
variational cluster approximations for the electron correlations.  
We have shown that both the  interband Coulomb interaction $U'$ and 
the electron-phonon coupling $\lambda$ tend to stabilize an excitonic CDW 
state. While at $\lambda=0$ the excitonic insulator exhibits an acoustic phase mode, 
any finite $\lambda$ fixes the phase of the order parameter and therefore eliminates
such a gapless excitation related to supertransport properties. 
The CDW typifies a predominantly excitonic and phononic state for small and large
ratios $8\lambda/U'$, respectively.  The Hund's rule coupling $J$, on the other hand, promotes
an excitonic SDW phase, which is further stabilized by pair-hopping processes, which also
fixes the phase of the order parameter. These results obtained for a generic microscopic model  Hamiltonian 
should contribute to a better understanding of exciton condensation in several material classes with 
strong electronic correlations. 

\begin{acknowledgments}
The authors would like to thank H. Beck, K. Seki, and H. Watanabe
for enlightening discussions. T.K.~acknowledges support from a JSPS Research 
Fellowship for Young Scientists.  This work was funded in part 
by a KAKENHI Grant No.~26400349 of Japan.  
B.Z.~and H.F.~are supported by Deutsche Forschungsgemeinschaft (Germany) through the Collaborative Research Center 652, Project B5.
\end{acknowledgments}


\begin{thebibliography}{99}
\bibitem{al} The use of the relation $U'=U-2J$, which is valid in the atomic limit, 
does not change the essential features of our results: 
see Ref.~\onlinecite{kaneko14} for details.
\end{thebibliography}
\end{document}